\documentclass[pre,showpacs,twocolumns]{revtex4}

\usepackage{amsmath} 
\usepackage{amssymb} 
\usepackage{graphicx}
\usepackage{color}

\newcommand{\be}{\begin{equation}} 
\newcommand{\ee}{\end{equation}}
\newcommand{\bea}{\begin{eqnarray}}   
\newcommand{\eea}{\end{eqnarray}}

\newcommand{\muone}{\mu_1} 
\newcommand{\mutwo}{\mu_2}

\newcommand{\butwo}{\mu^{(S)}_2}

\newcommand{\gtilde}{\gamma}

\begin{document}

\title{Non-equilibrium fluctuations in a driven stochastic Lorentz gas} 
 
\author{G. Gradenigo}
\affiliation{CNR-ISC and Dipartimento di Fisica, Universit\`a Sapienza, p.le A. Moro 2, 00185 Roma, Italy}
\author{U. Marini Bettolo Marconi} 
\affiliation{Scuola di Scienze e Tecnologie, Universit\`a  di Camerino, Camerino, Italy}
\author{A. Puglisi}
\affiliation{CNR-ISC and Dipartimento di Fisica, Universit\`a Sapienza, p.le A. Moro 2, 00185 Roma, Italy}
\author{A. Sarracino}
\affiliation{CNR-ISC and Dipartimento di Fisica, Universit\`a Sapienza, p.le A. Moro 2, 00185 Roma, Italy}

\begin{abstract}

  We study the stationary state of a one-dimensional kinetic
  model where a probe particle is driven by an external field
  $\mathcal{E}$ and collides, elastically or inelastically, with a
  bath of particles at temperature $T$. We focus on the stationary
  distribution of the velocity of the particle, and of two estimates of the total entropy
  production $\Delta s_{tot}$. One is the entropy production of the
  medium $\Delta s_m$, which is equal to the energy exchanged with the
  scatterers, divided by a parameter $\theta$, coinciding with the
  particle temperature at $\mathcal{E}=0$. The other is the work $W$
  done by the external field, again rescaled by $\theta$. At small
  $\mathcal{E}$, a good collapse of the two distributions is found: in
  this case the two quantities also verify the Fluctuation Relation
  (FR), indicating that both are good approximations of $\Delta
  s_{tot}$.  Differently, for large values of $\mathcal{E}$, the
  fluctuations of $W$ violate the FR, while $\Delta s_m$ still
  verifies it.

\end{abstract}


\maketitle

\section{Introduction}

The program of statistical mechanics, consisting in deriving
macroscopic properties of a system from the elementary interactions of
its constituents, is far from being fulfilled in out-of-equilibrium
conditions. In particular, the dissipation of energy in such cases
prevents the use of general equilibrium results, and forces one to rely on
a case-by-case model-dependent description.

One of the few general results for systems far from equilibrium has
been established in a series of works, starting from the seminal
papers by Evans and coworkers~\cite{ECM}. It consists in a family of
relations, generally referred to as Fluctuation Relations (FR), which
are very similar in form, but concern different quantities (e.g. phase
space contraction rate, entropy production, heat, work, etc.) and/or
different dynamical regimes (e.g. transients, stationary states,
etc.), as well as different kinds of non-equilibrium systems, either
deterministic or stochastic (see~\cite{BPRV08} and references
therein).  

In the framework of stochastic dynamics, the {\em microscopic}
definition of the entropy production relies on the knowledge of the
path probabilities for the model~\cite{Kurchan,LS99,CM99,C99,HS01,SS05,AG07}.
The study of the fluctuations of this quantity plays a
  central role in the characterization of small
  systems~\cite{bustamante,ritort-1}.  The connection with
macroscopic quantities that are \emph{reasonably} related to the
thermodynamic concept of ``entropy produced by the system'' has to be
carefully investigated in each specific model, as for instance
in~\cite{FM04,AGCGJP07,SVGP10}. In particular, the simple recipe,
useful near equilibrium, where the macroscopic entropy production is
expressed as the work done by external forces divided by the
temperature~\cite{DEGM}, is hardly of use far from equilibrium, often
because it is not clear which parameter plays the role of
temperature~\cite{leticia}.  For instance, in granular gases, due to
the dissipative character of interactions, the kinetic temperature
$T_g$ of the microscopic constituents does not always have a
thermodynamic role~\cite{BBDLMP05}.

In order to address such issues, we investigate the dynamics of a
stochastic Lorentz-like model where a particle is interacting with
random scatterers and is subjected to an external force. Stochastic
Lorentz models have been previously studied, focusing on transport
properties, e.g. on normal or anomalous diffusion, for instance
in~\cite{vb82,BCV04,MG05,GKP06,KPDCS06,AK11}. Our model is based on
the following ingredients: 1) the presence of an external field
$\mathcal{E}$ accelerating the probe particle, 2) scatterers of finite
mass which are randomly and uniformly distributed in space and move
with random velocities as extracted from a thermal bath at temperature
$T$ (it is therefore more reasonable to call them ``bath particles''),
3) collisions which can also be inelastic (the system always reaches a
stationary state), 4) a uniform collision probability which is
inspired from the so-called Maxwell-molecules models~\cite{E81}, and
which helps in simplifying analytical calculations. Such ingredients
provide a system in a non-equilibrium stationary state (NESS), due to
the presence of a finite stationary current of particles.  We do not
consider any time-dependent experimental protocol neither any
transformation between different NESS, that is, we study the system at
fixed value of the external parameter (the external field). For each
value of the external field, we start our measurements when the system
has already reached the corresponding NESS. Therefore, in our case,
there is not excess heat due to transient dynamics between different
NESS and the total heat exchanged equals the power dissipated to
sustain the NESS.

We show that, in this model, the microscopic entropy production is
well approximated, at large times, by the energy exchanged with the
bath, divided by a ``temperature'' $\theta$ which is that measured for
zero field, $\mathcal{E}=0$. Such a temperature turns out
  to be different from the temperature of scatterers $T$, unless the
  collisions are elastic, and coincides with the one of the probe
particle in the unperturbed process, in agreement with what recently
found in~\cite{EWS11}; at small values of the field, a macroscopically
accessible quantity well approximates the entropy production, and that
is the work done by the external field, divided by the same
temperature $\theta$.

\section{The model}

We consider an ensemble of probe particles of mass $m$ endowed with
scalar velocity $v$. Each probe particle only interacts with particles
of mass $M$ and velocity $V$ extracted from an equilibrium bath at
temperature $T$: such scatterers are distributed randomly and
uniformly in space and can hit the particle only once. The last
condition is important to guarantee unbounded motion and molecular
chaos even in one dimension, and can be thought as the effect of bath
particles moving in two (or more) dimensions, while the probe particle
can only move along a one-dimensional track.  Inspired from
Maxwell-molecules models~\cite{E81} and to their inelastic
generalization~\cite{BK00,BMP02,EB02}, we assume that the scattering
probability does not depend on the relative velocity of
colliders. Velocity of the particle changes from $v$ to $v'$ at each
collision, according to the rule:
\begin{equation} 
v'= \gamma v + (1- \gamma) V, 
\label{collrule}
\end{equation}
where 
\begin{equation}
\gamma=\frac{\zeta-\alpha}{1+\zeta} 
\end{equation}
with $\zeta=m/M$, and $\alpha$ is the coefficient of restitution
determining if the collision is elastic $\alpha=1$ or inelastic
$\alpha \in [0,1)$.  The velocity $V$ of the bath particles is a random
  variable generated from a Gaussian distribution with zero mean and
  variance $T/M$:
\begin{equation} 
P_S(V)=\sqrt{\frac{M}{2 \pi T }} \exp\left(-\frac{M}{2 T} V^2\right).
\label{scatt}
\end{equation}
In addition, the probe particle is accelerated by a uniform force
field $m\mathcal{E}$.  The resulting system can be assimilated to a
Lorentz gas model, where free flights in external field are
interrupted by random collisions with scatterers.

The model is summarized by the linear Boltzmann equation for the evolution of the velocity distribution of the probe particle
\begin{eqnarray}
\label{prob}
&&\tau_c \partial_t  P(v,t)+ \tau_c  \mathcal{E} \partial_v P(v,t) = \nonumber \\
&&-P(v,t)+ \frac{1}{1-\gamma}\int\!\!\!du\,
P(u,t)
P_S\left(\frac{ v-\gamma u} {1-\gamma}\right), 
\end{eqnarray}
where $\tau_c$ is the mean collision time.  In the following, we shall
compare analytical predictions with numerical simulations of
Eq.~(\ref{prob}). This equation, restricted to the
  particular case $\alpha=1$ and $m=M$ (that is $\gamma=0$), has been
  recently studied in~\cite{AP10}.


\section{Entropy production}

One of the most interesting features peculiar to out-of-equilibrium
stationary dynamics is that a finite rate of entropy production can be
measured in the system.  Microscopically, such a quantity is related to
the violation of detailed balance and gives a measure of how the
probability of observing a forward trajectory differs from the
probability of observing the time-reversed one~\cite{Kurchan,LS99}.
From the macroscopic point of view, the entropy production is related
to the presence of currents going through the system, due to spatial
gradients or due to the action of external driving
forces~\cite{DEGM,S76}. In simple examples~\cite{LS99,A06}, a bridge
between the two points of view can be verified, where the microscopic
entropy production turns out to be proportional to the product of a
flux by a force. The constant prefactor, in driven systems in contact
with a reservoir, is often found to be the bath temperature. However
this cannot be the general situation: for instance, for arbitrarily
strong external fields, the temperature of the system (e.g. the
kinetic one) may be far from that of the bath, or in extreme cases,
cannot even be defined.

The stochastic process considered here consists of two
parts: a deterministic evolution, due to the action of the external
field, plus a random contribution, due to the collisions with the
scatterers. Here, the deterministic process does not contribute to the entropy
production in the system, since the probability of a free fall for the
particles is symmetric under time-reversal.  It is then convenient to
rewrite Eq.~(\ref{prob}) as a Master Equation where the
transition rates describing the stochastic collisions between
particles explicitly appear
\begin{eqnarray} 
&&\tau_c\partial_tP(v,t)+\tau_c\mathcal{E}\partial_vP(v,t)= \nonumber \\
&&\int_{-\infty}^\infty dv' w(v|v')P(v',t) - \int_{-\infty}^\infty
dv' w(v'|v)P(v,t),  \nonumber \\
\label{Boltzmann1}
\end{eqnarray} 
with
\begin{equation}
\label{rate}
w(v'|v)=
\frac{1}{1-\gamma}\frac{1}{\sqrt{2\pi T/M}}
\exp\left\{
-\frac{M}{2T}\left[\frac{v'+\gamma v}{1-\gamma}\right]^2\right\}.
\end{equation}
To obtain this result we have put in Eq.~\eqref{prob} the form~(\ref{scatt}) for the
distribution of the scattering particles (see Ref~\cite{PVTW06} for the
general case, with different interaction kernels and arbitrary
dimension).  

For the process described by Eq.~(\ref{Boltzmann1}),
we can explicitly write the probability density $P[\{v(t)\}|v_0]$ of observing the
trajectory $\{v(s)\}_{s=t_0}^t$ in the interval $[t_0,t]$, with initial
and final values $v(t_0)=v_0$ and $v(t)=v_t$, respectively. The total
entropy production associated with each trajectory is defined as 
\begin{equation}
\Delta s_{tot}(t)=\log\frac{P[\{v(t)\}|v_0]P(v_0)}{P[\{\tilde{v}(t)\}|\tilde{v}_0]P(\tilde{v}_0)}=
\Delta s_m(t)+\log\frac{P(v_0)}{P(-v_t)},
\label{entropytot}
\end{equation}
with $\{\tilde{v}(s)=-v(t-s)\}_{s=t_0}^t$ the time-reversed path,
$P(v)$ the stationary distribution and $\Delta s_m$ the entropy
production of the medium~\cite{seifert05}, where as medium we refer to
the ensemble of scatterers.  Notice that in the reversed protocol the
electric field does not change the sign,
$\tilde{\mathcal{E}}=\mathcal{E}$.  Along a trajectory
where $N_c$ collisions occur at times $t_j$ (with $j\in [1,N_c]$) and
velocities change from $v_j=v(t_j^-)$ to $v_j'=v(t_j^+)$ one has
\begin{equation}
\frac{P[\{v(t)\}|v_0]}{P[\{\tilde{v}(t)\}|\tilde{v}_0]}=
\frac{\prod_{j=1}^{N_c}P_{nc}(\Delta t_j)w(v'_j|v_j)}
{\prod_{j=1}^{N_c}P_{nc}(\Delta t_j)w(-v_{j}|-v'_j)},
\label{ratio}
\end{equation}
where $P_{nc}(\Delta t_j)$ is the probability that no collision occurs
in the time interval $\Delta t_j=t_j-t_{j-1}$.  Since in the Maxwell
model the collision probability is independent of the particle
velocity, the time intervals $\Delta t_j$ between successive
collisions are distributed according to a Poissonian process with
$P_{nc}(\Delta t_j)\sim \exp(-\Delta t_j/\tau_c)$. In the ratio
between probabilities appearing in Eq.~(\ref{ratio}), the
contributions due to free flights in the backward trajectory exactly
cancel those coming from the forward trajectory. It is interesting to
mention that such cancellation is not a specific feature of the chosen
uniform collision probability: it can be verified also for other kinds of
interactions, e.g. for hard spheres.

The stochastic entropy production of the medium is then expressed in
terms of single event contributions $\delta s_m(v',v)$, where the
particle changes its velocity from $v$, before the collision, to $v'$,
after the collision,
\begin{equation} \label{smtot}
\Delta s_m(t)=\sum_{j=1}^{N_c(t)}\delta s_m(v'_j,v_j).
\end{equation}
Here $N_c(t)$ is the number of collisions occurred up to time $t$ and,
from Eq.~(\ref{rate}),
\begin{eqnarray} 
\delta s_m(v'_j,v_j)&=&\log\frac{w(v'_j|v_j)}{w(-v_j|-v'_j)} \nonumber \\ &=& 
 -\frac{M (1+\gamma)}{2T(1-\gamma)}(v'^2_j-v_j^2)  \nonumber \\ &=& - \frac{\delta E_{coll} (v'_j,v_j)}{\theta},
\label{DSDEcoll}
\end{eqnarray}
where $ \delta E_{coll} (v'_j,v_j)=m/2 (v'^2_j-v_j^2)$ is the energy received 
from the scatterers in a collision and 
\begin{equation}
\theta=T \zeta \frac{1-\gtilde}{1+\gtilde}= T\zeta\frac{1+\alpha}{1+2\zeta-\alpha}
\label{theta}
\end{equation}
is the kinetic temperature of the probe particle in zero field, 
as demonstrated below (see Eq.~(\ref{kineticT})).

For the average rate of entropy production in a stationary state we can write
\begin{eqnarray}
\langle \dot{s}_m(t)\rangle &=& \lim_{t\rightarrow \infty} \frac{\langle \Delta s_m(t)\rangle}{t} =
\frac{1}{\tau_c}\left\langle\delta s_m(v',v)\right\rangle \nonumber \\
&=&-\frac{1}{\tau_c}
\frac{M(1+\gamma)}{2T(1-\gamma)}(\langle v'^2\rangle_{post}- \langle v^2\rangle), \nonumber \\  
\label{entropy}
\end{eqnarray} 
where $\langle\cdots\rangle_{post}$ denotes the average over the
distribution of post-collision velocities. In order to compute
explicitly the average quantities appearing in Eq.~(\ref{entropy}),
we need to solve the Boltzmann equation~(\ref{prob}) in the
stationary limit.

\section{Stationary solution}

\begin{figure}[!t]
\includegraphics[width=.6\columnwidth,clip=true]{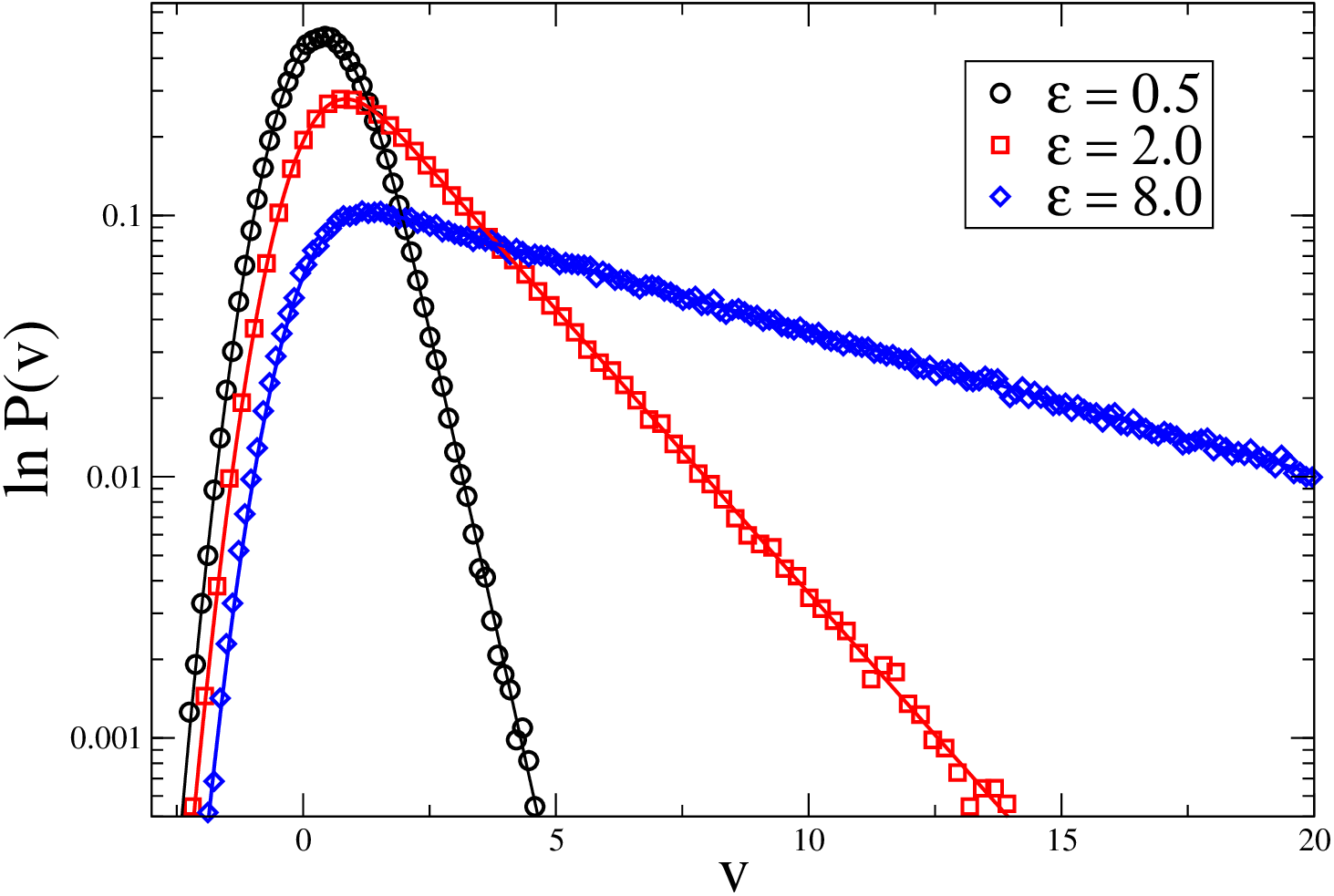}
\caption{(Color online) Comparison between the theoretical prediction of Eq.~(\ref{pdfgamma0})
(solid lines) and the numerical simulations (symbols)
for the stationary distribution in the case $\gamma=0$, $\tau_c=1$, $\zeta=0.5$, $T=1$
and for different values of the external field $\mathcal{E}=0.5,2,8$.}
\label{pdf_E}
\end{figure}

Defining the Fourier transform of the probability density
$\hat P(k,t)=\int_{-\infty}^{\infty} dv e^{ikv} P(v,t)$
and similarly for $\hat P_S(k)$,
the integral equation~(\ref{prob}) in Fourier space in the stationary 
limit assumes the convenient form:
\be
\label{foua}
 -ik \mathcal{E}\tau_c \hat P(k)  =-\hat P(k)+  \hat P(\gamma k) \hat P_S((1-\gamma)k). 
\ee
Let us start by considering the simplest case $\gamma=0$.  This
corresponds to $\zeta=\alpha$ and it has been considered
in~\cite{AP10,GP86}, for elastic particles with $m=M$.  Since $\hat
P(0)=1$, from Eq.~(\ref{foua}) we have
\begin{equation} \label{resum}
\hat P(k) = \frac{1}{1-i \mathcal{E}\tau_c k}\hat P_S(k) 
  = \frac{1}{1-i \mathcal{E}\tau_c k}\exp\left(-\frac{T}{2  M} k^2\right).
\end{equation}
The inversion of the Fourier transform  gives the sought distribution 
under the form of a convolution:
\be
  P(v)= b\sqrt{\frac{a}{\pi}}  \int_{0}^\infty  du \exp( -  a(v-u)^2 ) \times 
 \exp(-b u),
\label{pdfgamma0}
\ee
with $b=\frac{1}{\tau_c \mathcal{E}}$ and $a=\frac{M}{2T}$.  The result of the
convolution is that the tails for large $v$ are exponential and for
negative $v$ are Gaussian.
In the limit of infinite mass of the scatterers, $M\to \infty$, the distribution becomes one sided and reads:
\be
  P(v)= \theta(v) \frac{1}{\tau_c \mathcal{E}} \exp\left(-\frac{v}{\tau_c \mathcal{E}}\right).
\ee
In Fig.~\ref{pdf_E} the analytical prediction of Eq.~(\ref{pdfgamma0})
is compared with the pdf obtained in numerical simulations (see below).

For arbitrary values of $\gamma$, equation~(\ref{foua}) can be easily solved 
in the case of zero field $\mathcal{E}=0$, where one has
\be 
\hat P_0(k)=\exp\left(-\frac{\theta}{2 m}k^2\right), 
\ee 
with a temperature $\theta$.  Notice that $P_0(v)$
is a Maxwellian, but with a ``temperature'' $\theta$
which differs from $T$; one has $\theta=T$ only when $\alpha=1$.
In the presence of non-zero field,
no analogue of the closed 
formula \eqref{pdfgamma0} can be written. However, in this case one has access
to all the moments of the distribution. Indeed, assuming analyticity around $k=0$, 
for small enough $k$ one can write the expansion
\be
\hat P(k)  =\sum_{n=0}^\infty\frac{(ik)^n}{n!}\mu_n,
\label{series}
\ee 
where $\mu_n\equiv\langle v^n\rangle$.
Upon substituting expression~(\ref{series}) in Eq.~\eqref{foua}, and 
equating equal powers of $k$,
all the moments of the distribution can be obtained.
In particular, using Eq.~(\ref{scatt}), we have
\be
\mu_{1}=\frac{\tau_c \mathcal{E}}{1-\gamma}
\label{muone}
\ee
\be
\mutwo=\frac{2 \tau_c \mathcal{E}}{1-\gtilde^2}\muone+\frac{(1-\gtilde)^2}{1-\gtilde^2}\butwo.
\label{mutwo}
\ee 
It is interesting to note that the conductibility $\langle v
\rangle/\mathcal{E}=\tau_c/(1-\gamma)\equiv
\tau_c(1+\zeta)/(1+\alpha)$ {\em increases} when the system becomes
more inelastic ($\alpha$ reduced) or when the mass $M$ of the bath
particles is reduced.

We are now ready to compute the average entropy production.  
Indeed, substituting Eq.~\eqref{mutwo} in
Eq.~(\ref{entropy}), and using the fact that the distribution of
post-collisional velocities is by definition $\hat P_{post}(k)=\hat
P(\gamma k)\hat P_S((1-\gamma)k)$, we can write 
\begin{equation}
\langle \dot{s}_m(t) \rangle =\frac{(1+\gamma)}{(1-\gamma)^2}\frac{M}{T} \tau_c \mathcal{E}^2 \ge 0.
\label{entropy_ave}
\end{equation}
Expression~(\ref{entropy_ave}) can be related to the macroscopic
quantities present in the system, namely the external field $\mathcal{E}$ and
the current velocity $\langle v\rangle$.  Indeed, using Eq.~(\ref{theta}),
Eq.~(\ref{entropy_ave}) can be rewritten as
\begin{equation} 
\langle \dot{s}_m(t) \rangle 
= \frac{m}{\theta}\frac{\tau_c
  \mathcal{E}^2}{(1-\gamma)} = \frac{m \mathcal{E} \langle v \rangle}{\theta} \ge 0.  
\end{equation} 
Now let us consider the average work done by the external
field along a trajectory that spans the time interval $[0,t]$, 
$W(t) = \int_0^{t} F v(s) ds $, with $F=m \mathcal{E}$:
\begin{equation} 
\lim_{t\to\infty} \frac{1}{\theta}\frac{\langle W(t) \rangle}{t} = \frac{F \langle v\rangle}{\theta} = \langle \dot{s}_m(t) \rangle, 
\label{micromacro}
\end{equation}
i.e. the average \emph{macroscopic} work of the field divided by the
temperature $\theta$ corresponds to the average entropy production.
Notice that in Eq.~(\ref{micromacro}) the ``right'' temperature is
neither the bath temperature $T$, nor the kinetic temperature of the
probe particle in the presence of the field
\begin{equation}
T_g\equiv m \Bigl( \mu_2-\mu_1^2\Bigr)=\theta+ 
m \frac{(\tau_c \mathcal{E})^2}{(1-\gamma^2)},
\label{kineticT}
\end{equation}
but it is that of the unperturbed, not accelerated system. The
quantity $\theta$ represents an energy scale in the system, which
depends on the several parameters defining the model, namely
temperature of the scatterers $T$, mass ratio $\zeta$ and restitution
coefficient $\alpha$.  It is equal to the kinetic temperature of the
particle only in the absence of external field. In this case, if the
interactions are elastic, it equals the bath temperature $T$, as
expected. In general, for non zero field, the relation between $\theta$ and
the mean square velocity of the particle is expressed by Eq.~(\ref{kineticT}).


\section{Numerical simulations} 

In order to obtain numerically the stationary distribution $P(v)$
which solves Eq.~(\ref{prob}), we simulate the dynamics of a single
particle subject to a constant acceleration $\mathcal{E}$ and to
inelastic collisions with the scatterers, and average over $10^4$
realizations, for general $\gamma$.  Time is discretized in intervals $\delta t$, for a
total duration $\delta t N_t$. The particle is accelerated for $\delta
t$ under the influence of the constant field $\mathcal{E}$ and then a
collision with a scatterer is realized with probability
$p_{coll}=\delta t/\tau_c$. The distinguishing feature of Maxwell
molecules~\cite{E81} is that the probability of colliding is
independent of the velocity of the particle itself.  Then, the
particle is again uniformly accelerated by the field $\mathcal{E}$ and
the whole procedure is repeated. This ensures an average collision
time $\tau_c$, which is set as a parameter of the simulation.  For
each collision the velocity of the scatterers are extracted from a
constant Gaussian distribution, Eq.~(\ref{scatt}), and the collision
with the probe particle is realized with the inelastic rule,
Eq.~\eqref{collrule}.

In the numerical simulations we also studied the diffusive properties
of the system to find that the mean square displacement with respect
to the average motion, $d_2=\langle [x(t)-\langle x(t)
  \rangle]^2\rangle$, displays a ballistic behavior $d_2 \sim t^2$ on
short time scales with a crossover to a diffusional dynamics $d_2 \sim
t$ at large times. Such a ballistic-to-diffusive scenario, not
intuitive in the presence of an accelerating
field~\cite{BCV04,KPDCS06}, is consistent with the autocorrelation of
the particle's velocity which, up to numerical precision, decays as a
single exponential.

\begin{figure}[!t]
\includegraphics[width=.6\columnwidth,clip=true]{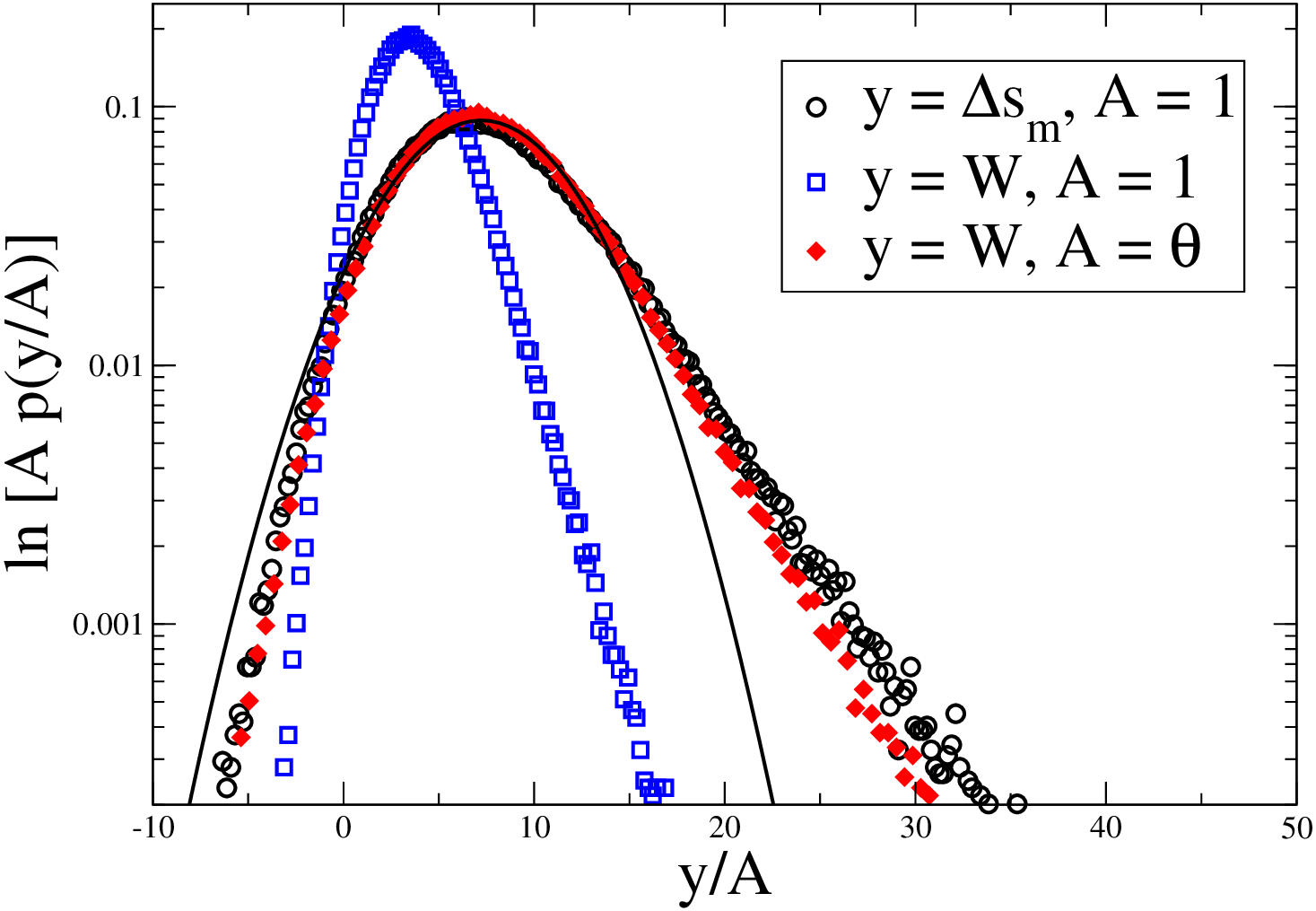}
\caption{(Color online) PDF of entropy production, work and rescaled
  work.  Parameters: $\gamma=0$, $\tau_c=1$, $\zeta=0.5$, $T=1$,
  $\mathcal{E}=0.5$ and $t=16\tau_c$.  The continuous line shows the
  Gaussian fit of $p(\Delta s_m)$.}
\label{pdf}
\end{figure}

\begin{figure}[!htb]
\includegraphics[width=.6\columnwidth,clip=true]{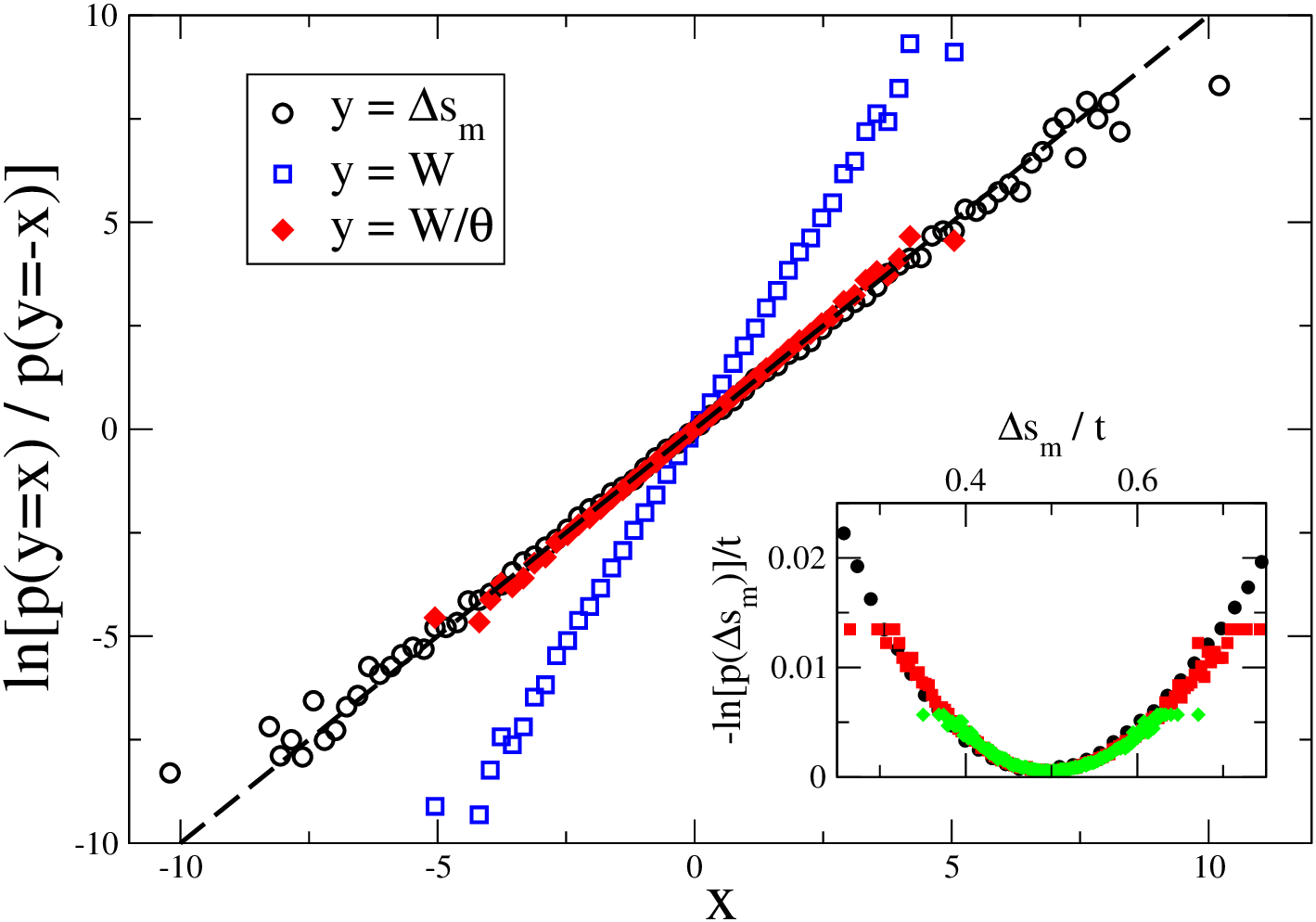}
\caption{(Color online) The symmetry relation~(\ref{FR}) obeyed by the
  entropy production $\Delta s_m(t)$ and by the rescaled work
  $W(t)/\theta$ obtained in numerical simulations with parameters:
  $\gamma=0$, $\tau_c=1$, $\zeta=0.5$, $T=1$, $\mathcal{E}=0.5$ and
  $t=16\tau_c$. Inset: collapse of the rate functions of the distribution
  of $\Delta s_m(t)$ for large times $t=128\tau_c$ (black circles),
  $t=512\tau_c$ (red squares) and $t=1024\tau_c$ (green diamond).}
\label{gcrelation}
\end{figure}


\section{The role of inelasticity}

A peculiarity of this model is that the inelastic collision rule can
be always mapped onto an elastic one: indeed we can always set
$\alpha=1$ and change the mass ratio $\zeta=m/M$ in order to keep
constant the parameter $\gamma$ which enters the collision rule.  In
practice, if also $m$ is kept fixed, we are changing only $M$ and thus
the width of the distribution in Eq.~(\ref{scatt}). Doing so we find no
relevant qualitative change in the physics of the system.  In
particular, let us notice that the entropy production in
Eq.~(\ref{entropy_ave}) depends only on $\gamma$ and vanishes with the
field $\mathcal{E}$, also with inelastic collisions.

The fact that, for the present model, the entropy production is not
affected by the inelasticity of collisions is in agreement with the
findings of~\cite{SVCP10}, where the dynamics of a single massive
intruder in a diluted granular gas of inelastic particles satisfying
the molecular chaos hypothesis was studied.  In that case the dynamics
of the probe particle was well described by a single linear Langevin
equation, so that the entropy production was zero by
definition. Without passing through a Langevin description, also the
Maxwell model presented here assumes the molecular chaos hypothesis for
the surrounding sea of scatterers. In both cases each collision of the
probe particle is performed with a velocity extracted from a constant
distribution independent of the previous history of the system:
despite the inelasticity of collisions, reversibility is guaranteed by
the molecular chaos hypothesis for the medium.  As it is clear from
Eq.~(\ref{entropy_ave}), in our model all irreversible effects are
generated by the external field.

There is only one special case in which the entropy production allows
us to distinguish the inelastic from the elastic interaction.  This is
the case of infinitely massive scatterers, namely $M \rightarrow
\infty $, which implies $\zeta=0$ and $\gamma=-\alpha$.  Keeping
finite the width of the velocity distribution of scatterers,
$T/M=\mu_2^{(S)}$, the average entropy production then reads
\begin{equation}  
\langle \dot{s}_m(t) \rangle = \frac{1-\alpha}{1+\alpha} \frac{\tau_c \mathcal{E}^2}{\mu_2^{(S)}},
\end{equation}
namely it vanishes in the elastic case.  This case is quite peculiar
because its stationary $P(v)$ has finite $\langle v \rangle$ but
infinite $\langle v^2 \rangle$~\cite{AP10}. For any other parameters
the model has finite or zero current and finite energy.

\section{Fluctuations of entropy production and work}

By definition the total entropy production~(\ref{entropytot}) 
must satisfy the FR for any value of $t$:
\begin{eqnarray}
\log\frac{p(\Delta s_{tot}(t)=x)}{p(\Delta s_{tot}(t)=-x)}=x,
\label{FR}
\end{eqnarray}
with $p(\Delta s_{tot}(t)=x)$ the probability density that in a time interval
$[0,t]$ the entropy produced by the system is $x$.  At large times
$t$, one usually neglects the term $\log[P(v_0)/P(-v_t)]$ which - in
the definition of $\Delta s_{tot}(t)$ - gives a contribution of order
$\mathcal{O}(1)$, and looks for the fulfillment of Eq.~\eqref{FR}
using $\Delta s_{tot}(t) \approx \Delta s_{m}(t)$, which should be the
leading part of order $\mathcal{O}(t)$. Our numerical simulations
suggest that this is the case also in this model, for any value of the
field $\mathcal{E}$.  As seen in Eqs.~\eqref{smtot}-\eqref{DSDEcoll},
$\Delta s_{m}(t)$ coincides with the energy that the probe particle
loses when colliding with the bath particles, divided by the kinetic
temperature of the particle itself measured at zero field.  In
Fig.~\ref{pdf} the distribution of $\Delta s_m(t)$ is plotted for a
certain value of $t$: it can be clearly seen from the superimposed fit
a remarkable deviation from gaussianity.

Inspired from the equality of the average value of the entropy and of
the work produced along a trajectory, see Eq.~(\ref{micromacro}), we
measured in numerical simulation also the fluctuations of the work
done by the external field on the particle, $W(t)$.  A remarkable
finding is that, for small values of $\mathcal{E}$, the PDF of the
work can be collapsed on the PDF of the entropy production by
exploiting the ``unperturbed'' temperature $\theta$ as a scaling
parameter.  In such cases, the measure of work fluctuations provides
an alternative way to measure the temperature of the unperturbed
system: indeed, as also verified in Fig.~\ref{gcrelation}, when
$W(t)/\theta$ satisfies the FR, one also has
$\log[p(W(t)=x)/p(W(t)=-x)]= x/\theta$.

\begin{figure}[!t]
\includegraphics[width=.6\columnwidth,clip=true]{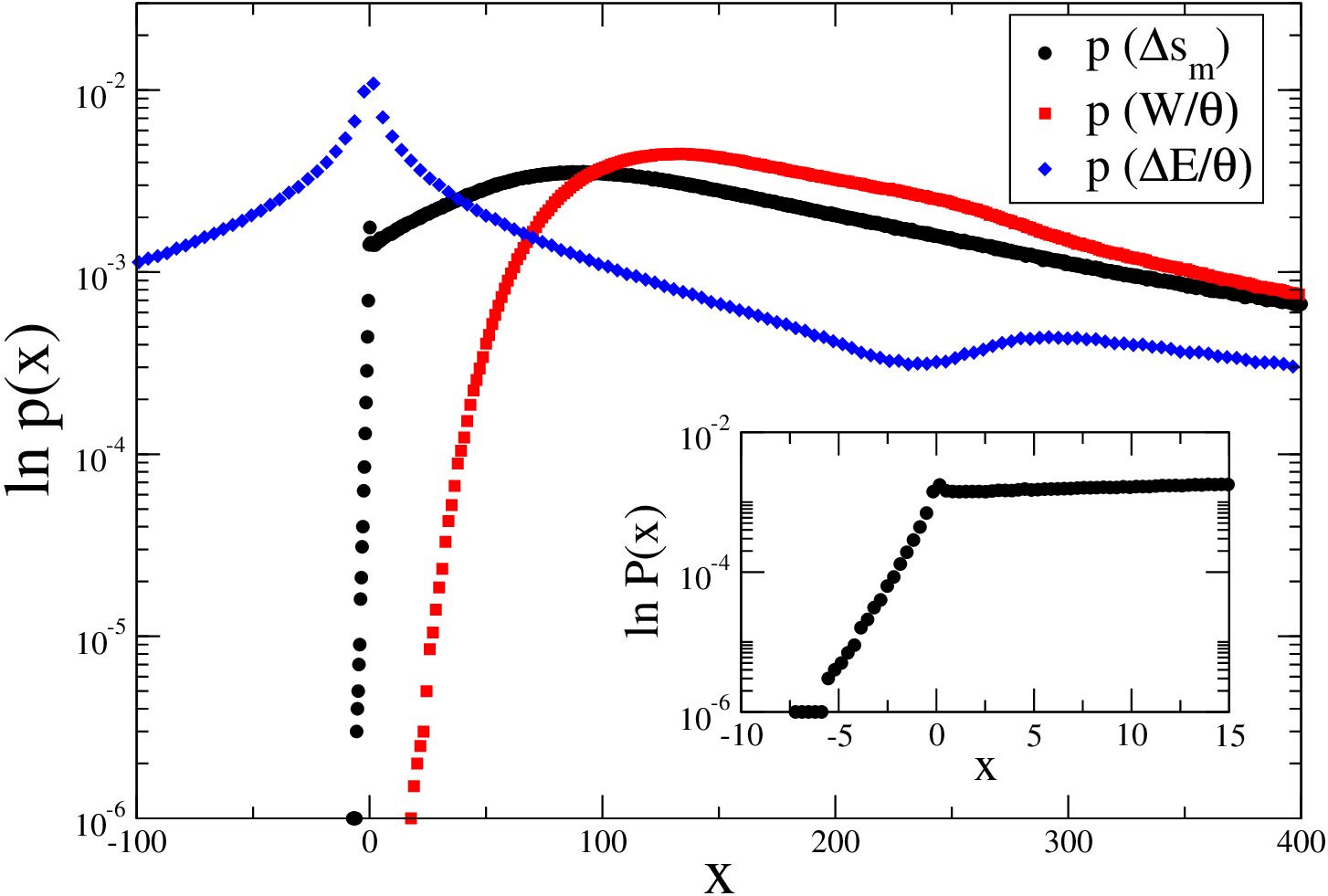}
\caption{(Color online) PDF of the quantities $\Delta s_m$, $W/\theta$ and
  $\Delta E/\theta$ measured in numerical simulations with
  parameters: $\gamma=0$, $\tau_c=1$, $\zeta=0.5$, $T=1$,
  $\mathcal{E}=8$ and $t=16\tau_c$.  Inset: zoom of the region around
  $x=0$ for the probability distribution of $\Delta s_m$.}
\label{strongE}
\end{figure}

By increasing the value of $\mathcal{E}$ while keeping fixed all the
other parameters of the simulation we find the following differences
with the small $\mathcal{E}$ situation: a) the stationary PDF of
velocity has a larger exponential tail and a very asymmetric shape; b)
the PDF of $\Delta s_m(t)$ also becomes very asymmetric and close to 0
has a shape very different from a Gaussian one, see inset of
Fig.~\ref{strongE}; c) the PDF of $W(t)$ cannot be collapsed on the
PDF of the entropy production with a simple rescaling; d) $\Delta
s_m(t)$ still verifies the FR, while $W(t)/\theta$ does not.

The breaking of the FR symmetry by the fluctuations of work at large
values of $\mathcal{E}$ can be interpreted considering the balance equations of
the energy absorbed and lost by the probe particle within a time
window. In particular, we have that $W(t) + \Delta E_{coll}(t)= \Delta
E(t)$, with $\Delta E(t)=\frac{m}{2}[v^2(t)-v^2(0)]$ and $\Delta
E_{coll}(t)=\sum_{j=1}^{N_c(t)}\delta E_{coll}(v'_j,v_j)$ and therefore,
exploiting the relation~(\ref{DSDEcoll}), we can write:
\begin{equation}
\Delta s_m(t) = \frac{1}{\theta} \left[ W(t) - \Delta E(t) \right],
\end{equation}
which puts in evidence that discrepancies between the distribution of
$\Delta s_m$ and $W/\theta$ are due to the ``boundary term'' $\Delta
E$.  Although $\Delta E(t)$ is of order $\mathcal{O}(1)$, it is known
to be dangerous for FR, even at large times, when its distribution has
exponential (or larger) tails~\cite{ZC03,PRV06}, as in our case (see
Fig.~\ref{strongE}).

In summary, we have discussed a non-equilibrium kinetic model, simple
enough to let most of the calculations accessible, such as the entropy
production or the moments of the stationary distribution, but still
displaying interesting properties, for instance strong non-Gaussian
behavior and non-trivial dependence on the external field. In
particular we have seen that the Fluctuation Relation is valid for the
distribution of the energy $\Delta E_{coll}$ lost in collisions with
the bath, and it also holds for the work done by the field, $W(t)$,
provided that $\mathcal{E}$ is low enough. For the fulfillment of the
FR, both quantities have to be divided by a temperature $\theta$ which
is different from the bath and the particle ones, but coincides with
the latter at $\mathcal{E}=0$.  We remark that such an observation is
highly non-trivial: it suggests that extreme care must be used when
entropy production is defined on phenomenological grounds, where one
is usually tempted to use more ``reasonable'' temperatures (e.g. the
bath or the system ones), forgetting the complexity of
far-from-equilibrium systems.

\begin{acknowledgments} 
  We warmly thank H.~Touchette, D.~Villamaina and A.~Vulpiani for a careful reading
  of the manuscript. The work of the authors is supported by the
  ``Granular-Chaos'' project, funded by the Italian MIUR under the
  FIRB-IDEAS grant number RBID08Z9JE.
\end{acknowledgments}


\bibliography{fluct}

\begin{thebibliography}{37}
\expandafter\ifx\csname natexlab\endcsname\relax\def\natexlab#1{#1}\fi
\expandafter\ifx\csname bibnamefont\endcsname\relax
  \def\bibnamefont#1{#1}\fi
\expandafter\ifx\csname bibfnamefont\endcsname\relax
  \def\bibfnamefont#1{#1}\fi
\expandafter\ifx\csname citenamefont\endcsname\relax
  \def\citenamefont#1{#1}\fi
\expandafter\ifx\csname url\endcsname\relax
  \def\url#1{\texttt{#1}}\fi
\expandafter\ifx\csname urlprefix\endcsname\relax\def\urlprefix{URL }\fi
\providecommand{\bibinfo}[2]{#2}
\providecommand{\eprint}[2][]{\url{#2}}

\bibitem[{\citenamefont{Evans et~al.}(1993)\citenamefont{Evans, Cohen, and
  Morriss}}]{ECM}
\bibinfo{author}{\bibfnamefont{D.~J.} \bibnamefont{Evans}},
  \bibinfo{author}{\bibfnamefont{E.~G.~D.} \bibnamefont{Cohen}},
  \bibnamefont{and} \bibinfo{author}{\bibfnamefont{G.~P.}
  \bibnamefont{Morriss}}, \bibinfo{journal}{Phys. Rev. Lett.}
  \textbf{\bibinfo{volume}{71}}, \bibinfo{pages}{2401} (\bibinfo{year}{1993}).

\bibitem[{\citenamefont{Marconi et~al.}(2008)\citenamefont{Marconi, Puglisi,
  Rondoni, and Vulpiani}}]{BPRV08}
\bibinfo{author}{\bibfnamefont{U.~M.~B.} \bibnamefont{Marconi}},
  \bibinfo{author}{\bibfnamefont{A.}~\bibnamefont{Puglisi}},
  \bibinfo{author}{\bibfnamefont{L.}~\bibnamefont{Rondoni}}, \bibnamefont{and}
  \bibinfo{author}{\bibfnamefont{A.}~\bibnamefont{Vulpiani}},
  \bibinfo{journal}{Phys. Rep.} \textbf{\bibinfo{volume}{461}},
  \bibinfo{pages}{111} (\bibinfo{year}{2008}).

\bibitem[{\citenamefont{Kurchan}(1998)}]{Kurchan}
\bibinfo{author}{\bibfnamefont{J.}~\bibnamefont{Kurchan}}, \bibinfo{journal}{J.
  Phys. A} \textbf{\bibinfo{volume}{31}}, \bibinfo{pages}{3719}
  (\bibinfo{year}{1998}).

\bibitem[{\citenamefont{Lebowitz and Spohn}(1999)}]{LS99}
\bibinfo{author}{\bibfnamefont{J.~L.} \bibnamefont{Lebowitz}} \bibnamefont{and}
  \bibinfo{author}{\bibfnamefont{H.}~\bibnamefont{Spohn}}, \bibinfo{journal}{J.
  Stat. Phys.} \textbf{\bibinfo{volume}{95}}, \bibinfo{pages}{333}
  (\bibinfo{year}{1999}).

\bibitem[{\citenamefont{Maes}(1999)}]{CM99}
\bibinfo{author}{\bibfnamefont{C.}~\bibnamefont{Maes}}, \bibinfo{journal}{J.
  Stat. Phys.} \textbf{\bibinfo{volume}{95}}, \bibinfo{pages}{367}
  (\bibinfo{year}{1999}).

\bibitem[{\citenamefont{Crooks}(1999)}]{C99}
\bibinfo{author}{\bibfnamefont{G.~E.} \bibnamefont{Crooks}},
  \bibinfo{journal}{Phys. Rev. E} \textbf{\bibinfo{volume}{60}},
  \bibinfo{pages}{2721} (\bibinfo{year}{1999}).

\bibitem[{\citenamefont{Hatano and Sasa}(2001)}]{HS01}
\bibinfo{author}{\bibfnamefont{T.}~\bibnamefont{Hatano}} \bibnamefont{and}
  \bibinfo{author}{\bibfnamefont{S.}~\bibnamefont{Sasa}},
  \bibinfo{journal}{Phys. Rev. Lett.} \textbf{\bibinfo{volume}{86}},
  \bibinfo{pages}{3463} (\bibinfo{year}{2001}).

\bibitem[{\citenamefont{Speck and Seifert}(2005)}]{SS05}
\bibinfo{author}{\bibfnamefont{T.}~\bibnamefont{Speck}} \bibnamefont{and}
  \bibinfo{author}{\bibfnamefont{U.}~\bibnamefont{Seifert}},
  \bibinfo{journal}{J. Phys. A} \textbf{\bibinfo{volume}{38}},
  \bibinfo{pages}{L581} (\bibinfo{year}{2005}).

\bibitem[{\citenamefont{Andrieux and Gaspard}(2007)}]{AG07}
\bibinfo{author}{\bibfnamefont{D.}~\bibnamefont{Andrieux}} \bibnamefont{and}
  \bibinfo{author}{\bibfnamefont{P.}~\bibnamefont{Gaspard}},
  \bibinfo{journal}{J. Stat. Phys.} \textbf{\bibinfo{volume}{127}},
  \bibinfo{pages}{107} (\bibinfo{year}{2007}).

\bibitem[{\citenamefont{Bustamante et~al.}(2005)\citenamefont{Bustamante,
  Liphardt, and Ritort}}]{bustamante}
\bibinfo{author}{\bibfnamefont{C.}~\bibnamefont{Bustamante}},
  \bibinfo{author}{\bibfnamefont{J.}~\bibnamefont{Liphardt}}, \bibnamefont{and}
  \bibinfo{author}{\bibfnamefont{F.}~\bibnamefont{Ritort}},
  \bibinfo{journal}{Physics Today} \textbf{\bibinfo{volume}{58}},
  \bibinfo{pages}{43} (\bibinfo{year}{2005}).

\bibitem[{\citenamefont{Collin et~al.}(2005)\citenamefont{Collin, Ritort,
  Jarzynski, Smith, Tinoco, and Bustamante}}]{ritort-1}
\bibinfo{author}{\bibfnamefont{D.}~\bibnamefont{Collin}},
  \bibinfo{author}{\bibfnamefont{F.}~\bibnamefont{Ritort}},
  \bibinfo{author}{\bibfnamefont{C.}~\bibnamefont{Jarzynski}},
  \bibinfo{author}{\bibfnamefont{S.~B.} \bibnamefont{Smith}},
  \bibinfo{author}{\bibfnamefont{I.}~\bibnamefont{Tinoco}}, \bibnamefont{and}
  \bibinfo{author}{\bibfnamefont{C.}~\bibnamefont{Bustamante}},
  \bibinfo{journal}{Nature} \textbf{\bibinfo{volume}{437}},
  \bibinfo{pages}{231} (\bibinfo{year}{2005}).

\bibitem[{\citenamefont{Feitosa and Menon}(2004)}]{FM04}
\bibinfo{author}{\bibfnamefont{K.}~\bibnamefont{Feitosa}} \bibnamefont{and}
  \bibinfo{author}{\bibfnamefont{N.}~\bibnamefont{Menon}},
  \bibinfo{journal}{Phys. Rev. Lett.} \textbf{\bibinfo{volume}{92}},
  \bibinfo{pages}{164301} (\bibinfo{year}{2004}).

\bibitem[{\citenamefont{Andrieux et~al.}(2007)\citenamefont{Andrieux, Gaspard,
  Ciliberto, Garnier, Joubaud, and Petrosyan}}]{AGCGJP07}
\bibinfo{author}{\bibfnamefont{D.}~\bibnamefont{Andrieux}},
  \bibinfo{author}{\bibfnamefont{P.}~\bibnamefont{Gaspard}},
  \bibinfo{author}{\bibfnamefont{S.}~\bibnamefont{Ciliberto}},
  \bibinfo{author}{\bibfnamefont{N.}~\bibnamefont{Garnier}},
  \bibinfo{author}{\bibfnamefont{S.}~\bibnamefont{Joubaud}}, \bibnamefont{and}
  \bibinfo{author}{\bibfnamefont{A.}~\bibnamefont{Petrosyan}},
  \bibinfo{journal}{Phy. Rev. Lett.} \textbf{\bibinfo{volume}{98}},
  \bibinfo{pages}{150601} (\bibinfo{year}{2007}).

\bibitem[{\citenamefont{Sarracino
  et~al.}(2010{\natexlab{a}})\citenamefont{Sarracino, Villamaina, Gradenigo,
  and Puglisi}}]{SVGP10}
\bibinfo{author}{\bibfnamefont{A.}~\bibnamefont{Sarracino}},
  \bibinfo{author}{\bibfnamefont{D.}~\bibnamefont{Villamaina}},
  \bibinfo{author}{\bibfnamefont{G.}~\bibnamefont{Gradenigo}},
  \bibnamefont{and} \bibinfo{author}{\bibfnamefont{A.}~\bibnamefont{Puglisi}},
  \bibinfo{journal}{Europhys. Lett.} \textbf{\bibinfo{volume}{92}},
  \bibinfo{pages}{34001} (\bibinfo{year}{2010}{\natexlab{a}}).

\bibitem[{\citenamefont{de~Groot and Mazur}(1984)}]{DEGM}
\bibinfo{author}{\bibfnamefont{S.~R.} \bibnamefont{de~Groot}} \bibnamefont{and}
  \bibinfo{author}{\bibfnamefont{P.}~\bibnamefont{Mazur}},
  \emph{\bibinfo{title}{Non-equilibrium thermodynamics}}
  (\bibinfo{publisher}{Dover Publications}, \bibinfo{address}{New York},
  \bibinfo{year}{1984}).

\bibitem[{\citenamefont{Cugliandolo}(2011)}]{leticia}
\bibinfo{author}{\bibfnamefont{L.}~\bibnamefont{Cugliandolo}},
  \bibinfo{journal}{J. Phys. A} \textbf{\bibinfo{volume}{44}},
  \bibinfo{pages}{483001} (\bibinfo{year}{2011}).

\bibitem[{\citenamefont{Baldassarri et~al.}(2005)\citenamefont{Baldassarri,
  Barrat, D'Anna, Loreto, Mayor, and Puglisi}}]{BBDLMP05}
\bibinfo{author}{\bibfnamefont{A.}~\bibnamefont{Baldassarri}},
  \bibinfo{author}{\bibfnamefont{A.}~\bibnamefont{Barrat}},
  \bibinfo{author}{\bibfnamefont{G.}~\bibnamefont{D'Anna}},
  \bibinfo{author}{\bibfnamefont{V.}~\bibnamefont{Loreto}},
  \bibinfo{author}{\bibfnamefont{P.}~\bibnamefont{Mayor}}, \bibnamefont{and}
  \bibinfo{author}{\bibfnamefont{A.}~\bibnamefont{Puglisi}},
  \bibinfo{journal}{Journal of Physics: Condensed Matter}
  \textbf{\bibinfo{volume}{17}}, \bibinfo{pages}{S2405} (\bibinfo{year}{2005}).

\bibitem[{\citenamefont{van Beijeren}(1982)}]{vb82}
\bibinfo{author}{\bibfnamefont{H.}~\bibnamefont{van Beijeren}},
  \bibinfo{journal}{Rev. Mod. Phys.} \textbf{\bibinfo{volume}{54}},
  \bibinfo{pages}{195} (\bibinfo{year}{1982}).

\bibitem[{\citenamefont{Bouchet et~al.}(2004)\citenamefont{Bouchet, Cecconi,
  and Vulpiani}}]{BCV04}
\bibinfo{author}{\bibfnamefont{F.}~\bibnamefont{Bouchet}},
  \bibinfo{author}{\bibfnamefont{F.}~\bibnamefont{Cecconi}}, \bibnamefont{and}
  \bibinfo{author}{\bibfnamefont{A.}~\bibnamefont{Vulpiani}},
  \bibinfo{journal}{Phys. Rev. Lett.} \textbf{\bibinfo{volume}{92}},
  \bibinfo{pages}{040601} (\bibinfo{year}{2004}).

\bibitem[{\citenamefont{M\'aty\'as and Gaspard}(2005)}]{MG05}
\bibinfo{author}{\bibfnamefont{L.}~\bibnamefont{M\'aty\'as}} \bibnamefont{and}
  \bibinfo{author}{\bibfnamefont{P.}~\bibnamefont{Gaspard}},
  \bibinfo{journal}{Phys. Rev. E} \textbf{\bibinfo{volume}{71}},
  \bibinfo{pages}{036147} (\bibinfo{year}{2005}).

\bibitem[{\citenamefont{Giardin\`a et~al.}(2006)\citenamefont{Giardin\`a,
  Kurchan, and Peliti}}]{GKP06}
\bibinfo{author}{\bibfnamefont{C.}~\bibnamefont{Giardin\`a}},
  \bibinfo{author}{\bibfnamefont{J.}~\bibnamefont{Kurchan}}, \bibnamefont{and}
  \bibinfo{author}{\bibfnamefont{L.}~\bibnamefont{Peliti}},
  \bibinfo{journal}{Phys. Rev. Lett.} \textbf{\bibinfo{volume}{96}},
  \bibinfo{pages}{120603} (\bibinfo{year}{2006}).

\bibitem[{\citenamefont{Karlis et~al.}(2006)\citenamefont{Karlis, Papachristou,
  Diakonos, Constantoudis, and Schmelcher}}]{KPDCS06}
\bibinfo{author}{\bibfnamefont{A.~K.} \bibnamefont{Karlis}},
  \bibinfo{author}{\bibfnamefont{P.~K.} \bibnamefont{Papachristou}},
  \bibinfo{author}{\bibfnamefont{F.~K.} \bibnamefont{Diakonos}},
  \bibinfo{author}{\bibfnamefont{V.}~\bibnamefont{Constantoudis}},
  \bibnamefont{and}
  \bibinfo{author}{\bibfnamefont{P.}~\bibnamefont{Schmelcher}},
  \bibinfo{journal}{Phys. Rev. Lett.} \textbf{\bibinfo{volume}{97}},
  \bibinfo{pages}{194102} (\bibinfo{year}{2006}).

\bibitem[{\citenamefont{D'Alessio and Krapivsky}(2011)}]{AK11}
\bibinfo{author}{\bibfnamefont{L.}~\bibnamefont{D'Alessio}} \bibnamefont{and}
  \bibinfo{author}{\bibfnamefont{P.~L.} \bibnamefont{Krapivsky}},
  \bibinfo{journal}{Phys. Rev. E} \textbf{\bibinfo{volume}{83}},
  \bibinfo{pages}{011107} (\bibinfo{year}{2011}).

\bibitem[{\citenamefont{Ernst}(1981)}]{E81}
\bibinfo{author}{\bibfnamefont{M.~H.} \bibnamefont{Ernst}},
  \bibinfo{journal}{Phys. Rep.} \textbf{\bibinfo{volume}{78}},
  \bibinfo{pages}{1} (\bibinfo{year}{1981}).

\bibitem[{\citenamefont{Evans et~al.}(2011)\citenamefont{Evans, Williams, and
  Searles}}]{EWS11}
\bibinfo{author}{\bibfnamefont{D.~J.} \bibnamefont{Evans}},
  \bibinfo{author}{\bibfnamefont{S.~R.} \bibnamefont{Williams}},
  \bibnamefont{and} \bibinfo{author}{\bibfnamefont{D.~J.}
  \bibnamefont{Searles}}, \bibinfo{journal}{J. Chem. Phys.}
  \textbf{\bibinfo{volume}{134}}, \bibinfo{pages}{204113}
  (\bibinfo{year}{2011}).

\bibitem[{\citenamefont{Ben-Naim and Krapivsky}(2000)}]{BK00}
\bibinfo{author}{\bibfnamefont{E.}~\bibnamefont{Ben-Naim}} \bibnamefont{and}
  \bibinfo{author}{\bibfnamefont{P.~L.} \bibnamefont{Krapivsky}},
  \bibinfo{journal}{Phys. Rev. E} \textbf{\bibinfo{volume}{61}},
  \bibinfo{pages}{R5} (\bibinfo{year}{2000}).

\bibitem[{\citenamefont{Baldassarri et~al.}(2002)\citenamefont{Baldassarri,
  Marconi, and Puglisi}}]{BMP02}
\bibinfo{author}{\bibfnamefont{A.}~\bibnamefont{Baldassarri}},
  \bibinfo{author}{\bibfnamefont{U.~M.~B.} \bibnamefont{Marconi}},
  \bibnamefont{and} \bibinfo{author}{\bibfnamefont{A.}~\bibnamefont{Puglisi}},
  \bibinfo{journal}{Europhys. Lett.} \textbf{\bibinfo{volume}{58}},
  \bibinfo{pages}{14} (\bibinfo{year}{2002}).

\bibitem[{\citenamefont{Ernst and Brito}(2002)}]{EB02}
\bibinfo{author}{\bibfnamefont{M.~H.} \bibnamefont{Ernst}} \bibnamefont{and}
  \bibinfo{author}{\bibfnamefont{R.}~\bibnamefont{Brito}},
  \bibinfo{journal}{Europhys. Lett.} \textbf{\bibinfo{volume}{58}},
  \bibinfo{pages}{182} (\bibinfo{year}{2002}).

\bibitem[{\citenamefont{Alastuey and Piasecki}(2010)}]{AP10}
\bibinfo{author}{\bibfnamefont{A.}~\bibnamefont{Alastuey}} \bibnamefont{and}
  \bibinfo{author}{\bibfnamefont{J.}~\bibnamefont{Piasecki}},
  \bibinfo{journal}{J. Stat. Phys.} \textbf{\bibinfo{volume}{139}},
  \bibinfo{pages}{991} (\bibinfo{year}{2010}).

\bibitem[{\citenamefont{Schnakenberg}(1976)}]{S76}
\bibinfo{author}{\bibfnamefont{J.}~\bibnamefont{Schnakenberg}},
  \bibinfo{journal}{Rev. Mod. Phys.} \textbf{\bibinfo{volume}{48}},
  \bibinfo{pages}{571} (\bibinfo{year}{1976}).

\bibitem[{\citenamefont{Astumian}(2006)}]{A06}
\bibinfo{author}{\bibfnamefont{R.~D.} \bibnamefont{Astumian}},
  \bibinfo{journal}{Am. J. Phys} \textbf{\bibinfo{volume}{74}},
  \bibinfo{pages}{683} (\bibinfo{year}{2006}).

\bibitem[{\citenamefont{Puglisi
  et~al.}(2006{\natexlab{a}})\citenamefont{Puglisi, Visco, Trizac, and van
  Wijland}}]{PVTW06}
\bibinfo{author}{\bibfnamefont{A.}~\bibnamefont{Puglisi}},
  \bibinfo{author}{\bibfnamefont{P.}~\bibnamefont{Visco}},
  \bibinfo{author}{\bibfnamefont{E.}~\bibnamefont{Trizac}}, \bibnamefont{and}
  \bibinfo{author}{\bibfnamefont{F.}~\bibnamefont{van Wijland}},
  \bibinfo{journal}{Phys. Rev. E} \textbf{\bibinfo{volume}{73}},
  \bibinfo{pages}{021301} (\bibinfo{year}{2006}{\natexlab{a}}).

\bibitem[{\citenamefont{Seifert}(2005)}]{seifert05}
\bibinfo{author}{\bibfnamefont{U.}~\bibnamefont{Seifert}},
  \bibinfo{journal}{Phys. Rev. Lett.} \textbf{\bibinfo{volume}{95}},
  \bibinfo{pages}{040602} (\bibinfo{year}{2005}).

\bibitem[{\citenamefont{Gervois and Piasecki}(1986)}]{GP86}
\bibinfo{author}{\bibfnamefont{A.}~\bibnamefont{Gervois}} \bibnamefont{and}
  \bibinfo{author}{\bibfnamefont{J.}~\bibnamefont{Piasecki}},
  \bibinfo{journal}{J. Stat. Phys.} \textbf{\bibinfo{volume}{42}},
  \bibinfo{pages}{1091} (\bibinfo{year}{1986}).

\bibitem[{\citenamefont{Sarracino
  et~al.}(2010{\natexlab{b}})\citenamefont{Sarracino, Villamaina, Costantini,
  and Puglisi}}]{SVCP10}
\bibinfo{author}{\bibfnamefont{A.}~\bibnamefont{Sarracino}},
  \bibinfo{author}{\bibfnamefont{D.}~\bibnamefont{Villamaina}},
  \bibinfo{author}{\bibfnamefont{G.}~\bibnamefont{Costantini}},
  \bibnamefont{and} \bibinfo{author}{\bibfnamefont{A.}~\bibnamefont{Puglisi}},
  \bibinfo{journal}{J. Stat. Mech.} p. \bibinfo{pages}{P04013}
  (\bibinfo{year}{2010}{\natexlab{b}}).

\bibitem[{\citenamefont{van Zon and Cohen}(2003)}]{ZC03}
\bibinfo{author}{\bibfnamefont{R.}~\bibnamefont{van Zon}} \bibnamefont{and}
  \bibinfo{author}{\bibfnamefont{E.~G.~D.} \bibnamefont{Cohen}},
  \bibinfo{journal}{Phys. Rev. Lett.} \textbf{\bibinfo{volume}{91}},
  \bibinfo{pages}{110601} (\bibinfo{year}{2003}).

\bibitem[{\citenamefont{Puglisi
  et~al.}(2006{\natexlab{b}})\citenamefont{Puglisi, Rondoni, and
  Vulpiani}}]{PRV06}
\bibinfo{author}{\bibfnamefont{A.}~\bibnamefont{Puglisi}},
  \bibinfo{author}{\bibfnamefont{L.}~\bibnamefont{Rondoni}}, \bibnamefont{and}
  \bibinfo{author}{\bibfnamefont{A.}~\bibnamefont{Vulpiani}},
  \bibinfo{journal}{J. Stat. Mech.} p. \bibinfo{pages}{P08010}
  (\bibinfo{year}{2006}{\natexlab{b}}).

\end{thebibliography}

 \end{document}